\documentclass{article}
\usepackage{amsfonts}

\usepackage{graphicx}
\usepackage{amsmath}

\newtheorem{theorem}{Theorem}

\newenvironment{proof}[1][Proof]{\textbf{#1.} }{\ \rule{0.5em}{0.5em}}
\input{tcilatex}

\begin{document}

\title{Hamilton-Jacobi-Bellman equations for\\
Quantum Filtering and Control}
\author{J. Gough$^{a}$, V.P. Belavkin$^{b}$, and O.G. Smolyanov$^{c}$ \\
$^{a}$Nottingham-Trent University, $^{b}$Nottingham University,\\
$^{c}$Moscow State Univeristy}
\date{}
\maketitle

\begin{abstract}
We exploit the separation of the filtering and control aspects of quantum
feedback control to consider the optimal control as a classical stochastic
problem on the space of quantum states. We derive the corresponding
Hamilton-Jacobi-Bellman equations using the elementary arguments of
classical control theory and show that this is equivalent, in the
Stratonovich calculus, to a stochastic Hamilton-Pontryagin setup. We show
that, for cost functionals that are linear in the state, the theory yields
the traditional Bellman equations treated so far in quantum feedback. A
controlled qubit with a feedback is considered as example.
\end{abstract}

\section{Introduction}

When engineers set about to control a classical system, they can evoke the
celebrated \textit{Separation Theorem} which allows them to treat the
problem of estimating the state of the system (based on typically partial
observations) from the problem of how to optimally control the system
(through feedback of these observations into the system dynamics), see for
instance \cite{Davis}. Remarkably, as it was pointed out for the first time
in \cite{B83}, this is also true when trying to control the quantum world,
see also \cite{B88},\cite{B99},\cite{BoutenEdwardsBelavkin}. To begin with,
the very act of measurement itself never supplies anything but incomplete
information about the state of a system and, as is well known, alters the
state in process. However, provided we use a non-demolition principle \cite
{B88} when measuring the system over time, we can apply a filter scheme for
state estimation continuously in time. The general theory of the continuous
in time nondemolition measurements and filtering was developed by Belavkin
in \cite{B88},\cite{B90},\cite{B92},\cite{B92(a)}, however we will use here
its final result for a simple quantum diffusion model described by the
quantum state filtering equation with a single white noise innovation, see
e.g. \cite{B89},\cite{WM},\cite{GutaBoutenMaassen}. We should emphasize that
the continuous-time filtering theory for this case can be obtained\ as the
limit of a discrete-time measurements where nothing beyond the standard von
Neumann projection postulate is used \cite{GoughSobolev},\cite{GoughSob2}, 
\cite{SM}, \cite{ST}. Once the filtered dynamics is known, the of optimal
feedback control of the system can then be formulated as a distinct problem.
Modern experimental physics has opened up unprecedented opportunities to
manipulate the quantum world, and feedback control has been already been
successfully implemented for real physical systems \cite{AASDM},\cite{GSDM}.
Currently, these activities have attracted interest in the related
mathematical issues such as stability, observability, etc., \cite
{BoutenEdwardsBelavkin},\cite{vHSM},\cite{DHJM},\cite{James}.

The separation of the classical world from the quantum world is, in
practice, the most notoriously troublesome task faced in modern physics. At
the very heart of this issue is the very different meanings we attach to the
word \textit{state}. What we want to remark upon, and exploit, is the fact
that the separation of the control problem from the filtering gives us just
the required separation of classical from quantum features. By the quantum
state we mean the von Neumann density matrix which yields all the
(stochastic) information available about the system at the current time -
this we also take to be the state in the sense used in control engineering.
All the quantum features are contained in this state, and the filtering
equation it satisfies may then to be understood as classical stochastic
differential equation which just happens to have solutions that are von
Neumann density matrix valued stochastic processes. The ensuing problem of
determining optimal control may then be viewed as a classical problem,
albeit on the unfamiliar state space of von Neumann density matrices rather
than the Euclidean spaces to which we are usually accustomed. Once we get
used to this setting, the problem of dynamical programming, Bellman's
optimality principle, and so on, can be formulated in the same spirit as
before.

We shall consider optimization for cost functions that are non-linear
functionals of the state. Traditionally quantum control has been restricted
to linear functions where - given the physical meaning attached to a quantum
state - the cost functions are therefore expectations of certain
observables. In this situation, which we consider as a special case, we see
that the distinction between classical and quantum features may be blurred:
that is, the classical information about the measurement observations can be
incorporated as additional randomness into the quantum state. This is the
likely reason why the separation does not seem to have been taken up before.

\section{Notations}

The Hilbert space for our fixed quantum system will be a complex, separable
Hilbert space $\mathfrak{h}$ . We shall use the following spaces of
operators: 
\begin{equation*}
\begin{tabular}{ll}
$\mathcal{A}=\mathfrak{B}\left( \mathfrak{h}\right) $ & - the Banach algebra
of bounded operators on $\mathfrak{h}$; \\ 
$\mathcal{A}_{\star }=\mathfrak{I}\left( \mathfrak{h}\right) $ & - the
predual space of trace-class operators on $\mathfrak{h}$; \\ 
$\mathcal{S}=\mathfrak{S}\left( \mathfrak{h}\right) $ & - the positive,
unital trace operators (states) on $\mathfrak{h}$.
\end{tabular}
\end{equation*}
The space $\mathcal{A}_{\star }$ equipped\ with the trace norm $\left\Vert
\varrho \right\Vert _{1}=\mathrm{tr}\left\vert \varrho \right\vert $ is the
complex Banach space, the dual of which is identified with the algebra $%
\mathcal{A}$ with usual operator norm. The natural duality between the
spaces $\mathcal{A}_{\star }$ and $\mathcal{A}$ is indicated by 
\begin{equation}
\left\langle \varrho ,X\right\rangle :=\mathrm{tr}\left\{ \varrho X\right\} ,
\end{equation}
for each $\varrho \in \mathcal{A}_{\star },X\in \mathcal{A}$. The positive
elements of $\mathcal{A}_{\star }$ normalized as $\left\Vert \varrho
\right\Vert _{1}=1$ are called normal states, and the extremal elements $%
\varrho \in \mathcal{S}$ of the convex set $\mathcal{S}\subset \mathcal{A}%
_{\star }$ correspond to pure quantum states. The symmetric tensor power $%
\mathcal{A}_{sym}^{\otimes 2}=\mathcal{A}\otimes _{sym}\mathcal{A}$ of the
algebra $\mathcal{A}$ is the subalgebra of $\mathfrak{B}\left( \mathfrak{h}%
^{\otimes 2}\right) $ of all bounded operators on the Hilbert product space $%
\mathfrak{h}^{\otimes 2}=\mathfrak{h}\otimes \mathfrak{h}$, commuting with
the unitary involutive operator $S=S^{\dagger }$ of permutations $\eta
_{1}\otimes \eta _{2}\mapsto \eta _{2}\otimes \eta _{1}$ for any $\eta
_{i}\in \mathfrak{h}$.

A map $\mathcal{L}\left( \cdot \right) $ from $\mathcal{A}=\mathfrak{B}%
\left( \mathfrak{h}\right) $ to itself is said to be a Lindblad generator if
it takes the form 
\begin{eqnarray}
\mathcal{L}\left( X\right) &=&-i\left[ X,H\right] +\sum_{\alpha }\mathcal{L}%
_{R_{\alpha }}\left( X\right) ,  \label{generator} \\
\mathcal{L}_{R}\left( X\right) &=&R^{\dagger }XR-\frac{1}{2}R^{\dagger }RX-%
\frac{1}{2}XR^{\dagger }R  \label{lindblad}
\end{eqnarray}
with $H$ self-adjoint, the $R_{\alpha }\in \mathcal{A}$ and $\sum_{\alpha
}R_{\alpha }^{\dagger }R_{\alpha }$ (ultraweakly convergent \cite{partha}
for an infinite set $\left\{ R_{\alpha }\right\} $). The generator is
Hamiltonian if it just takes form $i\left[ H,\cdot \right] $. The preadjoint 
$\mathcal{L}^{\prime }=\mathcal{L}_{\star }$ of a generator $\mathcal{L}$ is
defined on the preadjoint space $\mathcal{A}_{\star }$ through the relation $%
\left\langle \mathcal{L}^{\prime }\left( \varrho \right) ,X\right\rangle
=\left\langle \varrho ,\mathcal{L}\left( X\right) \right\rangle $. We note
that Lindblad generators have the property $\mathcal{L}\left( I\right) =0$
corresponding \ to conservation of the identity operator $I\in \mathcal{A}$
or, equivalently, $\mathrm{tr}\left\{ \mathcal{L}^{\prime }\left( \varrho
\right) \right\} =0$ for all $\varrho \in \mathcal{A}_{\star }$.

In quantum control theory it is necessary to consider time-dependent
generators $\mathcal{L}\left( t\right) $, through an integrable time
dependence of the controlled Hamiltonian $H\left( t\right) $, or more
generally due to a square-integrable time dependence of the coupling
operators $R_{\alpha }\left( t\right) $. We will always assume that these
integrability conditions, corresponding to the existence of the unique
solution $\varrho \left( t\right) =P_{t}\left( t_{0},\varrho _{0}\right) $
to the quantum state Master equation 
\begin{equation}
\frac{d}{dt}\varrho \left( t\right) =\mathcal{L}^{\prime }\left( t,\varrho
\left( t\right) \right) \equiv v\left( t,\varrho \left( t\right) \right) ,
\label{master equation}
\end{equation}
for all for $t\geq t_{0}$, given an initial condition $\varrho \left(
t_{0}\right) =\varrho _{0}\in \mathcal{S}$, are fulfilled.

\bigskip

Let $\mathsf{F}=\mathsf{F}\left[ \cdot \right] $ be a (nonlinear) functional 
$\varrho \mapsto \mathsf{F}\left[ \varrho \right] $ on $\mathcal{S}$, then
we say it admits a (Fr\'{e}chet) derivative if there exists a $\mathcal{A}$%
-valued function $\nabla _{\varrho }\mathsf{F}\left[ \cdot \right] $ on $%
\mathcal{A}_{\star }$ such that 
\begin{equation}
\lim_{h\rightarrow 0}\frac{1}{h}\left\{ \mathsf{F}\left[ \cdot +h\tau \right]
-\mathsf{F}\left[ \cdot \right] \right\} =\left\langle \tau ,\nabla
_{\varrho }\mathsf{F}\left[ \cdot \right] \right\rangle ,  \label{nabla_Q}
\end{equation}
for each $\tau \in \mathcal{A}_{\star }$. In the same spirit, a Hessian $%
\nabla _{\varrho }^{\otimes 2}\equiv \nabla _{\varrho }\otimes \nabla
_{\varrho }$ can be defined as a mapping from the functionals on $\mathcal{S}
$ to the $\mathcal{A}_{sym}^{\otimes 2}:=\mathcal{A}\otimes _{sym}\mathcal{A}
$-valued functionals, via 
\begin{gather}
\lim_{h,h^{\prime }\rightarrow 0}\frac{1}{hh^{\prime }}\left\{ \mathsf{F}%
\left[ \cdot +h\tau +h^{\prime }\tau ^{\prime }\right] -\mathsf{F}\left[
\cdot +h\tau \right] -\mathsf{F}\left[ \cdot +h^{\prime }\tau ^{\prime }%
\right] +\mathsf{F}\left[ \cdot \right] \right\}  \notag \\
=\left\langle \tau \otimes \tau ^{\prime },\nabla _{\varrho }\otimes \nabla
_{\varrho }\mathsf{F}\left[ \cdot \right] \right\rangle .
\end{gather}
and we say that the functional is twice continuously differentiable whenever 
$\nabla _{\varrho }^{\otimes 2}\mathsf{F}\left[ \cdot \right] $ exists and
is continuous in the trace norm topology.

Likewise, a functional $f:X\mapsto f\left[ X\right] $ on $\mathcal{A}$ is
said to admit an $\mathcal{A}_{\star }$-derivative if there exists an $%
\mathcal{A}_{\star }$-valued function $\nabla _{X}f\left[ \cdot \right] $ on 
$\mathcal{A}$ such that 
\begin{equation}
\lim_{h\rightarrow 0}\frac{1}{h}\left\{ f\left[ \cdot +hA\right] -f\left[
\cdot \right] \right\} =\left\langle \nabla _{X}f\left[ \cdot \right]
,A\right\rangle
\end{equation}
for each $A\in \mathfrak{B}\left( \mathfrak{h}\right) $.

With the customary abuses of differential notation, we have for instance 
\begin{equation*}
\nabla _{\varrho }f\left( \left\langle \varrho ,X\right\rangle \right)
=f^{\prime }\left( \left\langle \varrho ,X\right\rangle \right) X,\quad
\nabla _{X}f\left( \left\langle \varrho ,X\right\rangle \right) =f^{\prime
}\left( \left\langle \varrho ,X\right\rangle \right) \varrho .
\end{equation*}
Typically, we shall use $\nabla _{\varrho }$ more often, and tend denote it
by $\delta $ (as "inverse" to the notation $\varrho $), leaving the simple
notation $\nabla $ for $\nabla _{X}$.

\section{Quantum Filtering Equation}

The state of an individual continuously measured quantum system does not
coincide with the solution $\varrho \left( t\right) $ of the deterministic
master equation (\ref{master equation}) but is a $\mathcal{S}$-valued
stochastic process $\varrho _{\bullet }\left( t\right) :\omega \mapsto
\varrho _{\omega }\left( t\right) $ which depends on the random measurement
output $\omega =\left\{ \omega \left( t\right) \right\} \in \Omega $ in a
causal manner. We take the output process to constitute a white noise, in
which case we may work with the innovations process which will be a Wiener
process $W\left( t\right) $ defined in the generalized sense by $\frac{d}{dt}%
W\left( t\right) =\omega \left( t\right) $ with $W\left( 0\right) =0$. The
Belavkin quantum filtering equation in this case is \cite{B89},\cite{B92(a)}%
, \cite{WM},\cite{GutaBoutenMaassen} 
\begin{equation}
d\varrho _{\bullet }\left( t\right) =w\left( t,u\left( t\right) ,\varrho
_{\bullet }\left( t\right) \right) \,dt+\sigma \left( \varrho _{\bullet
}\left( t\right) \right) \,dW\left( t\right)  \label{Belavkin equation}
\end{equation}
where $dW\left( t\right) =W\left( t+dt\right) -W\left( t\right) $, the time
coefficient is 
\begin{equation}
w\left( t,u,\varrho \right) =i\left[ \varrho ,H\left( t,u\right) \right] +%
\mathcal{L}_{R}^{\prime }\left( \varrho \right) +\mathcal{L}_{L}^{\prime
}\left( \varrho \right) ,  \label{w}
\end{equation}
with $\mathcal{L}_{L}^{\prime }\left( \varrho \right) $ of the form given \
\ 
\begin{equation*}
\mathcal{L}_{L}^{\prime }\left( \varrho \right) =L\varrho L^{\dagger }-\frac{%
1}{2}\varrho L^{\dagger }L-\frac{1}{2}L^{\dagger }L\varrho ,
\end{equation*}
and the fluctuation coefficient is 
\begin{equation}
\sigma \left( \varrho \right) =L\varrho +\varrho L^{\dagger }-\left\langle
\varrho ,L+L^{\dagger }\right\rangle \varrho .  \label{sigma}
\end{equation}

Here $L$ is a bounded operator describing the coupling of the system to the
measurement apparatus.

The time coefficient $w$ consists of three separate terms: The first term is
Hamiltonian and depends on a control parameter $u$ belonging to some
parameter space $\mathcal{U}$ which we must specify at each time; the second
term is the adjoint of a general Lindblad generator $\mathcal{L}_{R}$ and
describes the uncontrolled, typically dissipative, effect of the
environment; the final term is adjoint to the Lindblad generator $\mathcal{L}%
_{L}\left( X\right) $ which is related to the coupling operator $L$.

The maps $w$ and $\sigma $ are required to be Lipschitz continuous in all
their components: for $L$ constant and bounded, this will be automatic for
the $\varrho $-variable with the notion of trace norm topology. We remark
that $\mathrm{tr}\left\{ \sigma \left( \varrho \right) \right\} =0$ and, by
conservativity, $\mathrm{tr}\left\{ w\left( t,u,\varrho \right) \right\} =0$
for all $\varrho \in \mathcal{A}_{\star }$. This implies that the
normalization $\mathrm{tr}\left\{ \varrho \right\} $ is a conserved quantity 
$\mathrm{tr}\left\{ \varrho _{\bullet }\left( t\right) \right\} =\mathrm{tr}%
\left\{ \varrho \right\} $ under the stochastic evolution (\ref{Belavkin
equation}).

A choice of control function $\left\{ u\left( t\right) :t\in \left[
T_{1},T_{2}\right] \right\} $ is required before we can solve the filtering
equation (\ref{Belavkin equation}) on the time interval $\left[ T_{1},T_{2}%
\right] $ for given initial state at time $T_{1}$. From what we have said
above, this is required to be a $\mathcal{U}$-valued function which we take
to be continuous for the moment.

Let $\left\{ P_{r,\omega }\left( t,\varrho \right) :r\geq t,\omega \in
\Omega \right\} $ be the solution $\varrho _{\bullet }\left( r\right)
=P_{r,\bullet }\left( t,\varrho \right) $ to (\ref{Belavkin equation})
starting in state $\varrho _{\omega }\left( t\right) =\varrho $ at time $r=t$
for all $\omega \in \Omega $. This will be a Markov process in $\mathcal{S}$
(embedded in the Banach space $\mathcal{A}_{\star }$), see for instance \cite
{DaPratoZabczyk}, and we remark that, for twice continuously differentiable
functionals $\mathsf{F}$ on $\mathcal{A}_{\star }$, we will have 
\begin{equation*}
\lim_{h\rightarrow 0^{+}}\frac{1}{h}\left\{ \mathbb{E}\left[ \mathsf{F}\left[
P_{t+h,\bullet }\left( t,\varrho \right) \right] -\mathsf{F}\left[ \varrho %
\right] \right] \right\} =D\left( t,u,\varrho \right) \mathsf{F}\left[
\varrho \right] ,
\end{equation*}
where $D\left( t,u,\varrho \right) $ is the elliptic operator defined by 
\begin{equation}
D\left( t,u,\varrho \right) \cdot =\left\langle w\left( t,u,\varrho \right)
,\delta \cdot \right\rangle +\frac{1}{2}\left\langle \sigma \left( \varrho
\right) \otimes \sigma \left( \varrho \right) ,\left( \delta \otimes \delta
\right) \cdot \right\rangle .  \label{D1}
\end{equation}
For the classical analogue of stochastic flows on manifolds, see for
instance \cite{Bismut}.

\subsection{Stratonovich Version}

We convert to the Stratonovich picture \cite{Stratonovich} by means of the
identity \cite{C.W. Gardiner} 
\begin{equation*}
\sigma \left( \varrho _{\bullet }\right) \,dW=\sigma \left( \varrho
_{\bullet }\right) \circ dW-\frac{1}{2}d\sigma \left( \varrho _{\bullet
}\right) .dW
\end{equation*}
and from (\ref{sigma}) we get 
\begin{equation*}
d\sigma \left( \varrho _{\bullet }\right) =Ld\varrho _{\bullet }+d\varrho
_{\bullet }L^{\dagger }-\left\langle d\varrho _{\bullet },L+L^{\dagger
}\right\rangle \varrho _{\bullet }-\left\langle \varrho _{\bullet
},L+L^{\dagger }\right\rangle d\varrho _{\bullet }\ -\left\langle d\varrho
_{\bullet },L+L^{\dagger }\right\rangle d\varrho _{\bullet }.
\end{equation*}
After a little algebra, we obtain the Stratonovich form of the Belavkin
filtering equation: 
\begin{equation}
d\varrho _{\bullet }=v\left( t,u,\varrho _{\bullet }\right) \,dt+\sigma
\left( \varrho _{\bullet }\right) \circ dW  \label{Belavkin-Stratonovich}
\end{equation}
where, with $\sigma \equiv \sigma \left( \varrho \right) $, 
\begin{eqnarray}
v\left( t,u,\varrho \right) &=&w\left( t,u,\varrho \right) -\frac{1}{2}%
\left\{ L\sigma +\sigma L^{\dagger }-\left\langle \sigma ,L+L^{\dagger
}\right\rangle \varrho -\left\langle \varrho ,L+L^{\dagger }\right\rangle
\sigma \right\}  \notag \\
&=&i\left[ \varrho ,H\left( t,u\right) \right] +\mathcal{L}_{R}^{\prime
}\left( \varrho \right) +\left\{ K\left( \varrho \right) \,\varrho +\varrho
\,K\left( \varrho \right) ^{\dagger }+\mathsf{F}\left( \varrho \right)
\,\varrho \right\}  \notag \\
&&  \label{v}
\end{eqnarray}
where we introduce the operator-valued function 
\begin{equation}
K\left( \varrho \right) :=-\frac{1}{2}\left( L+L^{\dagger }\right)
L+\left\langle \varrho ,L+L^{\dagger }\right\rangle L
\end{equation}
and the scalar-valued function 
\begin{equation}
\mathsf{F}\left( \varrho \right) :=\frac{1}{2}\left\langle \varrho
,L^{2}+2L^{\dagger }L+L^{\dagger 2}\right\rangle -\left\langle \varrho
,L+L^{\dagger }\right\rangle ^{2}.
\end{equation}

We refer to $w$ in (\ref{w}) and $v$ in (\ref{v}) as the \textit{It\^{o} and
Stratonovich state velocities}, respectively. We note that the decoherent
component $L\varrho L^{\dagger }$ appearing in $\mathcal{L}_{L}^{\prime }$,
and present in $w\left( t,u,\varrho \right) $, is now absent in $v\left(
t,u,\varrho \right) $.

The elliptical operator $D\left( t,u,\varrho \right) $ can then be put into
H\"{o}rmander form as 
\begin{equation}
D\left( t,u,\varrho \right) \left( \cdot \right) :=\left\langle v\left(
t,u,\varrho \right) ,\delta \cdot \right\rangle +\frac{1}{2}\left\langle
\sigma \left( \varrho \right) ,\delta \left\langle \sigma \left( \varrho
\right) ,\delta \cdot \right\rangle \right\rangle ,  \label{D2}
\end{equation}
by using the equality (\ref{v}) in the definition (\ref{D1}).

\section{Optimal Control}

The cost for a control function $\left\{ u\left( r\right) \right\} $ over
any time-interval $\left[ t,T\right] $ is random, taken to have the integral
form 
\begin{equation}
\mathsf{J}_{\omega }\left[ \left\{ u\left( r\right) \right\} ;t,\varrho %
\right] =\int_{t}^{T}\mathsf{C}\left( r,u\left( r\right) ,\varrho _{\omega
}\left( r\right) \right) dr+\mathsf{S}\left( \varrho _{\omega }\left(
T\right) \right)  \label{J}
\end{equation}
where $\left\{ \varrho _{\bullet }\left( r\right) :r\in \left[ t,T\right]
\right\} $ is the solution to the filtering equation with initial condition $%
\varrho _{\bullet }\left( t\right) =\varrho $. We assume that the cost
density $\mathsf{C}$ and the terminal cost $\mathsf{S}$ will be continuously
differentiable in each of its arguments. In fact, due to the statistical
interpretation of quantum states, we should consider only the linear
dependence 
\begin{equation}
\mathsf{C}\left( r,u,\varrho \right) =\left\langle \varrho ,C\left(
r,u\right) \right\rangle ,\;\mathsf{S}\left( \varrho \right) =\left\langle
\varrho ,S\right\rangle  \label{costs}
\end{equation}
of $\mathsf{C}$ and $\mathsf{S}$ on the state $\varrho $ as it was already
suggested in \cite{B83},\cite{B88},\cite{B99}. We will explicitly consider
this case later, but for the moment we will not use the linearity of $%
\mathsf{C}$ and $\mathsf{S}$.

The feedback control $u\left( t\right) $ is to be considered a random
variable $u_{\omega }\left( t\right) $ adapted with respect to the
innovation process $W\left( t\right) $ and so we therefore consider the
problem of minimizing its average cost value with respect to $\left\{
u_{\bullet }\left( t\right) \right\} $. To this end, we define the optimal
average cost to be 
\begin{equation}
\mathsf{S}\left( t,\varrho \right) :=\inf_{\left\{ u_{\bullet }\left(
r\right) \right\} }\,\mathbb{E}\left[ \mathsf{J}_{\bullet }\left[ \left\{
u_{\bullet }\left( r\right) \right\} ;t,\varrho \right] \right] ,  \label{S}
\end{equation}
where the minimum is considered over all measurable adapted control
strategies $\left\{ u_{\bullet }\left( r\right) :r\geq t\right\} $. The aim
of feedback control theory is then to find an optimal control strategy $%
\left\{ u_{\bullet }^{\ast }\left( t\right) \right\} $ and evaluate $\mathsf{%
S}\left( t,\varrho \right) $ on a fixed time interval $\left[ t_{0},T\right] 
$. Obviously that the cost $\mathsf{S}\left( t,\varrho \right) $ of the
optimal feedback control is in general smaller then the minimum of $\,%
\mathbb{E}\left[ \mathsf{J}_{\bullet }\left[ \left\{ u\right\} ;t,\varrho %
\right] \right] $ over nonstochastic strategies $\left\{ u\left( r\right)
\right\} $ only, which gives the solution of the open loop (without
feedback) quantum control problem. In the case of the linear costs (\ref
{costs}) this open-loop problem is equivalent to the following quantum
deterministic optimization problem which can be tackled by the classical
theory of optimal deterministic control in the corresponding Banach spaces.

\subsection{Bellman \& Hamilton-Pontryagin Optimality}

Let us first consider nonstochastic quantum optimal control theory assuming
that the state $\varrho \left( t\right) \in \mathcal{S}$ obeys the master
equation (\ref{master equation}) where $v\left( t,\cdot \right) =\mathcal{L}%
^{\prime }\left( t,u\left( t\right) ,\cdot \right) $ is an adjoint of some
Lindblad generator $\mathcal{L}^{\prime }\left( t,u,\cdot \right) \equiv
v\left( t,u,\cdot \right) $ for each $t$ and $u$ with, say, the control
being exercised in the Hamiltonian component $i\left[ \cdot ,H\left(
t,u\right) \right] $ as before. The control strategy $\left\{ u\left(
t\right) \right\} $ will be here non-random, as will be any specific cost 
\textsf{$J$}$\left[ \left\{ u\right\} ;t_{0},\varrho _{0}\right] $. For
times $t<t+\varepsilon <T$, one has 
\begin{equation*}
\mathsf{S}\left( t,\varrho \right) =\inf_{\left\{ u\right\} }\,\left\{
\int_{t}^{t+\varepsilon }\mathsf{C}\left( r,\varrho \left( r\right) ,u\left(
r\right) \right) dr+\int_{t+\varepsilon }^{T}\mathsf{C}\left( r,\varrho
\left( r\right) ,u\left( r\right) \right) dr+\mathsf{S}\left( \varrho \left(
T\right) \right) \right\} .
\end{equation*}

Suppose that $\left\{ u^{\ast }\left( r\right) :r\in \left[ t,T\right]
\right\} $ is an optimal control when starting in state $\varrho $ at time $%
t $, and denote by $\left\{ P_{r}\left( t,\varrho \right) :r\in \left[ t,T%
\right] \right\} $ the corresponding state dynamics $\varrho ^{\ast }\left(
r\right) =P_{r}\left( t,\varrho \right) $, $P_{t}=\varrho $. Bellman's
optimality principle \cite{Bellman},\cite{Davis} observes that the control $%
\left\{ u^{\ast }\left( r\right) :r\in \left[ t+\varepsilon ,T\right]
\right\} $ will then be optimal when starting from $\varrho ^{\ast }\left(
t+\varepsilon \right) $ at the later time $t+\varepsilon $. It therefore
follows that 
\begin{equation*}
\mathsf{S}\left( t,\varrho \right) =\inf_{\left\{ u\left( r\right) \right\}
}\,\left\{ \int_{t}^{t+\varepsilon }\mathsf{C}\left( r,u\left( r\right)
,\varrho \left( r\right) \right) dr+\mathsf{S}\left( t+\varepsilon ,\varrho
\left( t+\varepsilon \right) \right) \right\} .
\end{equation*}
For $\varepsilon $ small we expect that $\varrho \left( t+\varepsilon
\right) =\varrho +v\left( t,u\left( t\right) ,\varrho \right) \varepsilon
+o\left( \varepsilon \right) $ and provided that $\mathsf{S}$ is
sufficiently smooth we may make the Taylor expansion 
\begin{equation}
\mathsf{S}\left( t+\varepsilon ,\varrho \left( t+\varepsilon \right) \right)
=\left[ 1+\varepsilon \frac{\partial }{\partial t}+\varepsilon \left\langle
v\left( t,u\left( t\right) ,\varrho \right) ,\delta \right\rangle \right] 
\mathsf{S}\left( t,\varrho \right) +o\left( \varepsilon \right) .  \label{S1}
\end{equation}
In addition, we approximate 
\begin{equation*}
\int_{t}^{t+\varepsilon }\mathsf{C}\left( r,u\left( r\right) ,\varrho \left(
r\right) \right) dr=\varepsilon \mathsf{C}\left( t,u\left( t\right) ,\varrho
\right) +o\left( \varepsilon \right)
\end{equation*}
and conclude that 
\begin{equation*}
\mathsf{S}\left( t,\varrho \right) =\inf_{u\in U}\,\left\{ \left[
1+\varepsilon \left( \mathsf{C}\left( t,u,\varrho \right) +\frac{\partial }{%
\partial t}+\left\langle v\left( t,u,\varrho \right) ,\delta \right\rangle
\right) \right] \mathsf{S}\left( t,\varrho \right) \right\} +o\left(
\varepsilon \right)
\end{equation*}
where now the infimum is taken over the point-value of $u\left( t\right)
=u\in U$. In the limit $\varepsilon \rightarrow 0$, one obtains the equation 
\begin{equation}
\frac{\partial }{\partial t}\mathsf{S}\left( t,\varrho \right) +\inf_{u\in 
\mathcal{U}}\left\{ \mathsf{C}\left( t,u,\varrho \right) +\left\langle
v\left( t,u,\varrho \right) ,\nabla \mathsf{S}\left( t,\varrho \right)
\right\rangle \right\} =0.  \label{HJB1}
\end{equation}
The equation is then to be solved subject to the terminal condition 
\begin{equation}
\mathsf{S}\left( T,\varrho \right) =\mathsf{S}\left( \varrho \right) .
\end{equation}

We may introduce the Pontryagin Hamiltonian function on $\left[ 0,T\right]
\times \mathcal{S}\times \mathcal{A}$ defined by the Legendre-Frenchel
transform 
\begin{equation}
\mathcal{H}_{v}\left( t,\varrho ,X\right) :=\sup_{u\in \mathcal{U}}\left\{
\left\langle v\left( t,u,\varrho \right) ,\lambda I-X\right\rangle -\mathsf{C%
}\left( t,u,\varrho \right) \right\} ,
\end{equation}
(which in fact does not depend on $\lambda \in \mathbb{C}$ since $%
\left\langle v\left( t,u,\varrho \right) ,I\right\rangle =0$). It should be
emphasized that these Hamiltonians are purely classical devices which may be
called super-Hamiltonians to be distinguished from $H$. We may then rewrite (%
\ref{HJB1}) as the (backward) \textit{Hamilton-Jacobi} equation 
\begin{equation}
\frac{\partial }{\partial t}\mathsf{S}\left( t,\varrho \right) =\mathcal{H}%
_{v}\left( t,\varrho ,\delta \mathsf{S}\left( t,\varrho \right) \right) .
\label{HJB2}
\end{equation}

The operator-valued function $X\left( t,\varrho \right) =\delta \mathsf{S}%
\left( t,\varrho \right) $ satisfying then the equation $\frac{d}{dt}%
X=\delta \mathcal{H}_{v}\left( t,\rho ,X\right) $ is referred to as the 
\textit{co-state}, with the terminal condition $X\left( T,\varrho \right)
=\delta \mathsf{S}\left( \varrho \right) $. We remark that, if $u^{\ast
}\left( t,\varrho ,X\right) $ is an optimal control minimizing 
\begin{equation*}
\mathcal{K}_{v}\left( t,u,\varrho ,X\right) =\left\langle v\left(
t,u,\varrho \right) ,\lambda I-X\right\rangle -\mathsf{C}\left( t,u,\varrho
\right) \text{,}
\end{equation*}
then the corresponding state dynamical equation $\frac{d}{dt}\varrho
=v\left( t,u^{\ast }\left( t,\varrho ,X\right) ,\varrho \right) $ in terms
of its optimal solution $P_{t}\equiv P_{t}\left( t_{0},\varrho _{0}\right) $
corresponding to $P_{t_{0}}=\varrho ^{\ast }\left( t_{0}\right) \equiv
\varrho _{0}$ can be written as $\dot{P}=-\nabla _{Q}\mathcal{H}_{v}\left(
t,P,Q\right) $ noting that 
\begin{equation*}
\mathcal{H}_{v}\left( t,P,Q\right) =\left\langle v\left( t,u^{\ast }\left(
t,P,Q\right) ,P\right) ,\lambda I-Q\right\rangle -\mathsf{C}\left( t,u^{\ast
}\left( t,P,Q\right) ,P\right) ,
\end{equation*}
where $Q_{t}=X\left( t\right) $ is the solution $X\left( t\right)
=Q_{t}\left( T,S\right) $ of $\dot{Q}=\nabla _{P}\mathcal{H}_{v}\left(
t,P,Q\right) $ corresponding to $Q_{T}=\delta \mathsf{S}\left( \varrho
\right) \equiv S$. Thus we may equivalently consider the system of
Hamiltonian equations 
\begin{equation}
\left\{ 
\begin{array}{c}
\dot{P}_{t}+\nabla _{Q}\mathcal{H}_{v}\left( t,P_{t},Q_{t}\right) =0, \\ 
\dot{Q}_{t}-\nabla _{P}\mathcal{H}_{v}\left( t,P_{t},Q_{t}\right) =0.
\end{array}
\right.  \label{Ham}
\end{equation}
which we refer to as the \textit{Hamilton-Pontryagin equations}, in direct
analogy with the classical case \cite{Pontryagin}. If we set $u^{\ast
}=u^{\ast }\left( t,P,Q\right) $ such that $\mathcal{K}_{v}\left( t,u^{\ast
},P,Q\right) =\sup_{u\in \mathcal{U}}\mathcal{K}_{v}\left( t,u,P,Q\right) $,
then the Pontryagin maximum principle is the observation that, for state and
co-state $\left\{ P\right\} $ and $\left\{ Q\right\} $ respectively leading
to optimality, we will have $\mathcal{K}_{v}\left( t,u,P,Q\right) \leq 
\mathcal{H}_{v}\left( t,P,Q\right) $ with equality for $u=u^{\ast }\left(
t,P,Q\right) $ maximizing $\mathcal{K}_{v}\left( t,u,P,Q\right) $.

\subsection{Bellman Equation for Filtered Dynamics}

We now consider the stochastic differential equation (\ref{Belavkin equation}%
) for the filtered state in place of the master equation (\ref{master
equation}). This time, the cost is random and we consider the problem of
computing the minimum average cost as in (\ref{S}). The Bellman principle
can however be applied once more. As before, we let $\left\{ u_{\omega
}^{\ast }\left( t\right) \right\} $ be a stochastic adapted control leading
to optimality and let $\varrho _{\omega }^{\ast }\left( r\right)
=P_{r,\omega }\left( t,\varrho \right) $ be the corresponding state
trajectory (now a stochastic process) starting from $\varrho $ at time $t$.
Again choosing $t<t+\varepsilon <T$, we have by the Bellman principle 
\begin{gather*}
\mathbb{E}\left[ \mathsf{S}\left( t+\varepsilon ,\varrho _{\bullet }^{\ast
}\left( t+\varepsilon \right) \right) \right] =\mathsf{S}\left( t,\varrho
\right) \\
+\inf_{u\in \mathcal{U}}\,\mathbb{E}\left\{ \frac{\partial \mathsf{S}}{%
\partial t}\left( t,\varrho \right) +\mathsf{C}\left( t,u,\varrho \right)
+D\left( t,u,\varrho \right) \mathsf{S}\left( t,\varrho \right) \right\}
\varepsilon +o\left( \varepsilon \right)
\end{gather*}
Taking the limit $\varepsilon \rightarrow 0$ yields the diffusive Bellman
equation 
\begin{equation*}
\dfrac{\partial \mathsf{S}}{\partial t}+\inf_{u\in \mathcal{U}}\left\{ 
\mathsf{C}\left( t,u,\varrho \right) +D\left( t,u,\varrho \right) \mathsf{S}%
\left( t,\varrho \right) \right\} =0.
\end{equation*}
This equation to be solved backward with the terminal condition $\mathsf{S}%
\left( T,\varrho \right) =\mathsf{S}\left( \varrho \right) $. Using the
super-Hamiltonian function 
\begin{equation*}
\mathcal{H}_{w}\left( t,\varrho ,X\right) :=\sup_{u\in \mathcal{U}}\left\{
\left\langle w\left( t,u,\varrho \right) ,\lambda I-X\right\rangle -\mathsf{C%
}\left( t,u,\varrho \right) \right\}
\end{equation*}
this can be written either in the Hamilton-Jacobi form as 
\begin{equation}
\frac{\partial \mathsf{S}}{\partial t}+\frac{1}{2}\left\langle \sigma \left(
\varrho \right) \otimes \sigma \left( \varrho \right) ,\left( \delta \otimes
\delta \right) \mathsf{S}\right\rangle =\mathcal{H}_{w}\left( t,\varrho
,\delta \mathsf{S}\right) .  \label{HJBs}
\end{equation}

\section{Stochastic Hamilton-Jacobi-Bellman Equation}

An alternative approach to deriving the equation (\ref{HJBs}) will now be
formulated. First of all we make a Wong-Zakai approximation \cite{Wong Zakai}
to the Stratonovich filtering equation (\ref{Belavkin-Stratonovich}). This
is achieved by introducing a differentiable process $W_{\omega }^{\left(
\lambda \right) }\left( t\right) =\int_{0}^{t}\omega ^{\left( \lambda
\right) }\left( r\right) dr$ converging to the Wiener noise $W_{\omega
}\left( t\right) $ as $\lambda \rightarrow 0$ almost surely and uniformly
for $t\in \left[ 0,T\right] $. We may then expect the same type of
convergence for $\left\{ \varrho _{\omega }^{\left( \lambda \right) }\left(
t\right) \right\} $, the solution to the random ODE 
\begin{equation*}
\frac{d}{dt}\varrho _{\omega }^{\left( \lambda \right) }\left( t\right)
=v\left( t,u\left( t\right) ,\varrho _{\omega }^{\left( \lambda \right)
}\left( t\right) \right) +\sigma \left( \varrho _{\omega }^{\left( \lambda
\right) }\left( t\right) \right) \omega ^{\left( \lambda \right) }\left(
t\right)
\end{equation*}
with non random initial condition $\,\varrho _{\omega }^{\left( \lambda
\right) }\left( t_{0}\right) =\varrho _{0}$, to the solution $\left\{
\varrho _{\omega }\left( t\right) \right\} $ with the same initial data $%
\varrho _{\omega }\left( t_{0}\right) =\varrho _{0}$.

If we fix the output $\omega \in \Omega $, then we have an equivalent
non-random dynamical system for which we will have a minimal cost function
and we denote this as $\mathsf{S}_{\omega }^{\left( \lambda \right) }\left(
t_{0},\varrho _{0}\right) $. Note that this depends on the assumed
realization of the measurement output process and on the approximation
parameter $\lambda $. The HJB equation for $\mathsf{S}_{\omega }^{\left(
\lambda \right) }\left( t,\varrho \right) $ will be (\ref{HJB2}) with $%
v\left( t\right) $ now replaced by $v\left( t\right) +\sigma \omega ^{\left(
\lambda \right) }\left( t\right) $: 
\begin{equation*}
\frac{\partial }{\partial t}\mathsf{S}_{\omega }^{\left( \lambda \right)
}+\left\langle \sigma \left( \varrho \right) ,\delta \mathsf{S}_{\omega
}^{\left( \lambda \right) }\right\rangle \,\omega ^{\left( \lambda \right)
}\left( t\right) =\mathcal{H}_{v}\left( t,\varrho ,\delta \mathsf{S}_{\omega
}^{\left( \lambda \right) }\right)
\end{equation*}
Since $\sigma \left( \varrho \right) \omega ^{\left( \lambda \right) }\left(
t\right) $ doesn't depend on $u$, the corresponding optimal strategy $%
u_{\omega }^{\ast }\left( t\right) $ as the solution of the optimization
problem 
\begin{equation*}
\inf_{u\in \mathcal{U}}\left\{ \mathsf{C}\left( u,\varrho \right)
+\left\langle v\left( u,\varrho \right) +\sigma \left( \varrho \right)
\omega ^{\left( \lambda \right) },X\right\rangle \right\} =\left\langle
\sigma \left( \varrho \right) \omega ^{\left( \lambda \right)
},X\right\rangle -\mathcal{H}_{v}\left( \varrho ,X\right)
\end{equation*}
is the same function $u^{\ast }\left( t,\varrho ,X\right) $ of $\varrho
=\varrho _{\omega }^{\left( \lambda \right) }\left( t\right) $ and $X=\delta 
\mathsf{S}_{\omega }^{\left( \lambda \right) }$, independent of $\omega
^{\left( \lambda \right) }\left( t\right) $. In the limit $\lambda
\rightarrow 0$ we obtain the Stratonovich SDE 
\begin{equation}
d\mathsf{S}_{\omega }\left( t,\varrho \right) +\left\langle \sigma \left(
\varrho \right) ,\delta \mathsf{S}_{\omega }\left( t,\varrho \right)
\right\rangle \circ dW\left( t\right) =\mathcal{H}_{v}\left( t,\varrho
,\delta \mathsf{S}_{\omega }\left( t,\varrho \right) \right) dt.
\label{SHJB}
\end{equation}
which may be called a stochastic Hamilton-Jacobi-Bellman equation.

\subsection{Interpretation of the Stochastic HJB equation}

The expression $\mathsf{S}_{\omega }\left( t_{0},\varrho _{0}\right) $ gives
the optimal cost from start time $t_{0}$ to terminal time $T$ when we begin
in state $\varrho _{0}$ and have measurement output $\omega \in \Omega $. It
evidently depends on the information $\{\omega \left( r\right) :r\in \left[
t_{0},T\right] \}$ only and is statistically independent of the noise $%
W\left( t\right) =\omega _{t}$ prior to time $t_{0}$. In this sense, the
stochastic action $\mathsf{S}_{\omega }\left( t,\varrho \right) $ is \textbf{%
backward-adapted}. This point is of crucial importance: it means that the
stochastic Hamiltonian-Jacobi-Bellman theory is not related directly to the
stochastic Hamilton-Jacobi theory \cite{TZ1}\ where both the state and the
action are always taken as be forward-adapted; it also means that we need to
be careful when converting (\ref{SHJB}) to It\^{o} form. This is a direct
consequence of the fact that Bellman's principle works by backward induction.

Let us introduce the following time-reversed notations 
\begin{equation*}
\tau :=T-t,\;\tilde{W}\left( \tau \right) :=W\left( T-\tau \right) =W\left(
t\right) \text{ and }\mathsf{\tilde{S}}_{\omega }\left( \tau ,\varrho
\right) :=\mathsf{S}_{\omega }\left( T-\tau ,\varrho \right) =\mathsf{S}%
_{\omega }\left( t,\varrho \right) .
\end{equation*}
The process $\tau \mapsto \mathsf{\tilde{S}}_{\bullet }\left( \tau ,\varrho
\right) $ is forward adapted to the filtration generated by $\tilde{W}$:
that is $\mathsf{\tilde{S}}_{\bullet }\left( \tau ,\varrho \right) $ is
measurable with respect to the sigma algebra generated by \{$\tilde{W}\left(
\sigma \right) $: $\sigma \in \left[ 0,\tau \right] \}$. Note that the
It\^{o} differential $d\tilde{W}\left( \tau \right) =\tilde{W}\left( \tau
+\varepsilon \right) -\tilde{W}\left( \tau \right) $ coincides with $W\left(
t-\varepsilon \right) -W\left( t\right) \equiv -\tilde{d}W\left( t\right) $
for $t=T-\tau \equiv \tilde{\tau}$.

\begin{theorem}
The stochastic process $\left\{ \mathsf{S}_{\bullet }\left( t,\varrho
\right) :t\in \left[ 0,T\right] \right\} $ satisfies the backward It\^{o}
SDE 
\begin{equation}
d\mathsf{S}_{\bullet }+\frac{1}{2}\left\langle \sigma ,\nabla \left\langle
\sigma ,\nabla \mathsf{S}_{\bullet }\right\rangle \right\rangle
dt+\left\langle \sigma ,\nabla \mathsf{S}_{\bullet }\right\rangle \;\tilde{d}%
W=\mathcal{H}_{v}\left( t,\varrho ,\nabla \mathsf{S}_{\bullet }\right) dt
\label{SHJB'}
\end{equation}
where $\tilde{d}W\left( t\right) :=W\left( t\right) -W\left( t-dt\right) $
is the past-pointing It\^{o} differential.
\end{theorem}

\begin{proof}
For simplicity, we suppress the $\varrho $ dependences. We shall take $%
\varepsilon >0$ to be infinitesimal and recast (\ref{SHJB}) in the form 
\begin{gather*}
\left[ \mathsf{S}_{\bullet }\left( t+\frac{1}{2}\varepsilon \right) -\mathsf{%
S}_{\bullet }\left( t-\frac{1}{2}\varepsilon \right) \right] -\mathcal{H}%
_{v}\left( t,\delta \mathsf{S}_{\bullet }\left( t\right) \right) \varepsilon
\\
+\left\langle \sigma ,\delta \mathsf{S}_{\bullet }\left( t\right)
\right\rangle \left[ W\left( t+\frac{1}{2}\varepsilon \right) -W\left( t-%
\frac{1}{2}\varepsilon \right) \right] =o\left( \varepsilon \right) .
\end{gather*}
In time-reversed notations, this becomes 
\begin{gather*}
\left[ \mathsf{\tilde{S}}_{\bullet }\left( \tau -\frac{1}{2}\varepsilon
\right) -\mathsf{\tilde{S}}_{\bullet }\left( \tau +\frac{1}{2}\varepsilon
\right) \right] -\mathcal{\tilde{H}}_{v}\left( t,-\delta \mathsf{\tilde{S}}%
_{\bullet }\left( \tau \right) \right) \varepsilon \\
+\left\langle \sigma ,\delta \mathsf{\tilde{S}}_{\bullet }\left( \tau
\right) \right\rangle \left[ \tilde{W}\left( \tau -\frac{1}{2}\varepsilon
\right) -\tilde{W}\left( \tau +\frac{1}{2}\varepsilon \right) \right]
=o\left( \varepsilon \right) ,
\end{gather*}
where $\mathcal{\tilde{H}}_{v}\left( t,\varrho ,X\right) =\mathcal{H}%
_{v}\left( t,\varrho ,-X\right) $. We then have the forward-time equation 
\begin{gather*}
\left[ \mathsf{\tilde{S}}_{\bullet }\left( \tau +\frac{1}{2}\varepsilon
\right) -\mathsf{\tilde{S}}_{\bullet }\left( \tau -\frac{1}{2}\varepsilon
\right) \right] +\mathcal{\tilde{H}}_{v}\left( t,-\delta \mathsf{\tilde{S}}%
_{\bullet }\left( \tau \right) \right) \varepsilon \\
+\left\langle \sigma ,\delta \mathsf{\tilde{S}}_{\bullet }\left( \tau
\right) \right\rangle \left[ \tilde{W}\left( \tau +\frac{1}{2}\varepsilon
\right) -\tilde{W}\left( \tau -\frac{1}{2}\varepsilon \right) \right]
=o\left( \varepsilon \right) ,
\end{gather*}
and using the It\^{o}-Stratonovich transformation 
\begin{eqnarray*}
&&\left\langle \sigma ,\delta \mathsf{\tilde{S}}\left( \tau \right)
\right\rangle \left[ \tilde{W}\left( \tau +\frac{1}{2}\varepsilon \right) -%
\tilde{W}\left( \tau -\frac{1}{2}\varepsilon \right) \right] +o\left(
\varepsilon \right) \\
&=&\left\langle \sigma ,\delta \mathsf{\tilde{S}}\left( \tau \right)
\right\rangle \left[ \tilde{W}\left( \tau +\varepsilon \right) -\tilde{W}%
\left( \tau \right) \right] -\frac{1}{2}\left\langle \sigma ,\delta
\left\langle \sigma ,\delta \mathsf{\tilde{S}}\left( \tau \right)
\right\rangle \right\rangle \varepsilon ,
\end{eqnarray*}
we get by substitution 
\begin{gather*}
\left[ \mathsf{\tilde{S}}\left( \tau +\frac{1}{2}\varepsilon \right) -%
\mathsf{\tilde{S}}\left( \tau -\frac{1}{2}\varepsilon \right) \right] -\frac{%
1}{2}\left\langle \sigma ,\delta \left\langle \sigma ,\delta \mathsf{\tilde{S%
}}\left( \tau \right) \right\rangle \right\rangle \varepsilon \\
+\mathcal{\tilde{H}}_{v}\left( t,-\delta \mathsf{\tilde{S}}\left( \tau
\right) \right) \varepsilon +\left\langle \sigma ,\delta \mathsf{\tilde{S}}%
\left( \tau \right) \right\rangle \left[ \tilde{W}\left( \tau +\varepsilon
\right) -\tilde{W}\left( \tau \right) \right] =o\left( \varepsilon \right) .
\end{gather*}
or, in the backward form for the original $\mathsf{S}_{\bullet }\left(
t,\varrho \right) =\mathsf{\tilde{S}}_{\bullet }\left( T-t,\varrho \right) $%
, 
\begin{gather*}
\left[ \mathsf{S}\left( t+\frac{1}{2}\varepsilon \right) -\mathsf{S}\left( t-%
\frac{1}{2}\varepsilon \right) \right] +\frac{1}{2}\left\langle \sigma
,\delta \left\langle \sigma ,\delta \mathsf{S}\left( t\right) \right\rangle
\right\rangle \varepsilon \\
-\mathcal{H}_{v}\left( t,\delta \mathsf{S}\left( t\right) \right)
\varepsilon +\left\langle \sigma ,\delta \mathsf{S}\left( t\right)
\right\rangle \left[ W\left( t\right) -W\left( t-\varepsilon \right) \right]
=o\left( \varepsilon \right) .
\end{gather*}
In the differential form this clearly is the same as (\ref{SHJB'}).
\end{proof}

\bigskip

If we denote by $\mathbb{E}^{\left( t_{0},\varrho _{0}\right) }$ expectation
(conditional on $\varrho _{\omega }\left( t_{0}\right) =\varrho _{0}$), then 
$\mathbb{E}^{\left( t_{0},\varrho _{0}\right) }\left[ \left\langle \sigma
,\delta \mathsf{S}_{\bullet }\left( t\right) \right\rangle \;\tilde{d}%
W\left( t\right) \right] =0$ since $\mathsf{S}_{\bullet }\left( t\right) $,
and its derivatives, are independent of the mean-zero past-point It\^{o}
differentials. We then have as a corollary that the averaged cost $\mathsf{S}%
\left( t,\varrho \right) $ defined by 
\begin{equation*}
\mathsf{S}\left( t_{0},\varrho _{0}\right) :=\mathbb{E}^{\left(
t_{0},\varrho _{0}\right) }\left[ \mathsf{S}_{\bullet }\left( t_{0},\varrho
_{0}\right) \right]
\end{equation*}
will satisfy the equivalent diffusive Hamilton-Jacobi equation 
\begin{equation}
\frac{\partial \mathsf{S}}{\partial t}+\frac{1}{2}\left\langle \sigma
,\delta \left\langle \sigma ,\delta \mathsf{S}\right\rangle \right\rangle =%
\mathcal{H}_{v}\left( t,\varrho ,\delta \mathsf{S}\right)  \label{HJB}
\end{equation}
which is the H\"{o}rmander form of the Bellman equation for optimal cost $%
\mathsf{S}\left( t,\varrho \right) $.

\section{Linear-State Cost}

A special case is applied to quantum mechanics when $\mathsf{C}\left(
t,u,\varrho \right) $ and $\mathsf{S}\left( \varrho \right) $ are both
linear (\ref{costs}) in the state $\varrho $ with quadratic dependence of $%
\mathsf{C}$ on $u$. Let us specify for simplicity to a time-independent cost
observable with control parameter $u=\left( u^{1},\cdots ,u^{n}\right) \in 
\mathbb{R}^{n}$ and having a quadratic dependence of the form (Einstein
index notation!) 
\begin{equation*}
C\left( u\right) =\frac{1}{2}g_{\alpha \beta }u^{\alpha }u^{\beta
}+u^{\alpha }F_{\alpha }+C_{0}
\end{equation*}
where $\left( g_{\alpha \beta }\right) $ are the components of a symmetric
positive definite metric with inverse denoted $\left( g^{\alpha \beta
}\right) $ and $F_{1},\cdots ,F_{n},C_{0}$ are fixed bounded operators. We
take control Hamiltonian operator to be 
\begin{equation*}
H\left( u\right) =u^{\alpha }V_{\alpha }
\end{equation*}
where $V_{1},\cdots ,V_{n}$ are fixed bounded observables. Our aim is to
find the optimal value $u^{\ast }$ for each pair $\left( P,Q\right) $ giving
a minimum to $\left\langle P,C\left( u\right) \right\rangle +\left\langle
w\left( t,u,P\right) ,Q\right\rangle =-\mathcal{K}_{w}\left( t,u,P,Q\right) $%
: we will have 
\begin{eqnarray*}
0 &=&\frac{\partial }{\partial u^{\alpha }}\left\{ \left\langle P,C\left(
u\right) \right\rangle +\left\langle w\left( t,u,P\right) ,Q\right\rangle
\right\} \\
&=&g_{\alpha \beta }u^{\beta }+\left\langle P,F_{\alpha }\right\rangle
+\left\langle i\left[ P,V_{\alpha }\right] ,Q\right\rangle .
\end{eqnarray*}
Thus the optimal control $u^{\ast }\left( P,Q\right) $ is given by the
components 
\begin{equation*}
u^{\alpha }=-g^{\alpha \beta }\left\langle P,F_{\beta }+\frac{1}{i}\left[
Q,V_{\beta }\right] \right\rangle .
\end{equation*}
This yields a unique point of infimum and on substituting we determine that 
\begin{eqnarray*}
\mathcal{H}_{w}\left( P,Q\right) &=&\frac{1}{2}g^{\alpha \beta }\left\langle
P,F_{\alpha }+\frac{1}{i}\left[ Q,V_{\alpha }\right] \right\rangle
\left\langle P,F_{\beta }+\frac{1}{i}\left[ Q,V_{\beta }\right] \right\rangle
\\
&&-\left\langle P,C_{0}+\mathcal{L}_{R}\left( Q\right) +\mathcal{L}%
_{L}\left( Q\right) \right\rangle .
\end{eqnarray*}
As a result, the Hamilton-Jacobi-Bellman equation takes the form 
\begin{gather*}
\frac{\partial \mathsf{S}}{\partial t}-\frac{1}{2}g^{\alpha \beta
}\left\langle \varrho ,F_{\alpha }+\frac{1}{i}\left[ \delta \mathsf{S}%
,V_{\alpha }\right] \right\rangle \left\langle \varrho ,F_{\beta }+\frac{1}{i%
}\left[ \delta \mathsf{S},V_{\beta }\right] \right\rangle \\
+\left\langle \varrho ,C_{0}+\mathcal{L}_{R}\left( \delta \mathsf{S}\right) +%
\mathcal{L}_{L}\left( \delta \mathsf{S}\right) \right\rangle +\frac{1}{2}%
\left\langle \sigma \left( \varrho \right) \otimes \sigma \left( \varrho
\right) ,\left( \delta \otimes \delta \right) \mathsf{S}\right\rangle =0.
\end{gather*}
The terminal condition being that $\mathsf{S}\left( \varrho ,T\right)
=\left\langle \varrho ,S\right\rangle $.

\subsection{Controlled Qubit}

Let us illustrate the above for the case of a qubit (two-state system). The
problem we consider is similar to the one formulated in \cite
{BoutenEdwardsBelavkin}. Denoting the Pauli spin vector by $\vec{\varsigma}%
=\left( \varsigma _{x},\varsigma _{y},\varsigma _{z}\right) $ with 
\begin{equation*}
\varsigma _{x}=\left( 
\begin{array}{cc}
0 & 1 \\ 
1 & 0
\end{array}
\right) ,\varsigma _{y}=\left( 
\begin{array}{cc}
0 & -i \\ 
i & 0
\end{array}
\right) ,\varsigma _{z}=\left( 
\begin{array}{cc}
1 & 0 \\ 
0 & -1
\end{array}
\right) ,
\end{equation*}
we may represent each state by polarization vector $\vec{r}\in \mathbb{R}%
^{3} $ as 
\begin{equation*}
\varrho =\frac{1}{2}\left( 1+\vec{p}.\vec{\varsigma}\right)
\end{equation*}
where $\left\vert \vec{p}\right\vert \leq 1$, while any observable takes the
form 
\begin{equation*}
Q=q_{0}+\vec{q}.\vec{\varsigma}
\end{equation*}
and we have the duality $\left\langle \varrho ,Q\right\rangle =q_{0}+\vec{q}.%
\vec{p}$. We shall write $\vec{p}=\left( x,y,z\right) $ and $\vec{q}=\left(
q_{x},q_{y},p_{z}\right) $.

Let us suppose that we have maximal control of the Hamilton component of the
dynamics, that is, we set 
\begin{equation*}
H\left( \vec{u}\right) =\frac{1}{2}\vec{u}.\vec{\varsigma}
\end{equation*}
with control variable $\vec{u}\in \mathbb{R}^{3}$. We also ignore the effect
of the environment and take $\mathcal{L}_{R}\equiv 0$. For simplicity, we
shall take the cost to have the form 
\begin{equation*}
\mathsf{C}\left( t,u,\varrho \right) =\frac{1}{2}\left\vert \vec{u}%
\,\right\vert ^{2}
\end{equation*}
and we take the coupling of the system to the measurement apparatus to be
determined by the operator 
\begin{equation*}
L=\frac{1}{2}\kappa \varsigma _{z}.
\end{equation*}

Explicitly we have 
\begin{equation*}
\left\langle w\left( t,u,\varrho \right) ,Q\right\rangle =\vec{u}.\left( 
\vec{p}\times \vec{q}\right) -\frac{1}{2}\kappa \left( xq_{x}+yq_{y}\right)
\end{equation*}
from which we see that the minimizing control is $\vec{u}^{\ast }=\vec{q}%
\times \vec{p}$ leading to the Hamiltonian function 
\begin{equation*}
\mathcal{H}_{w}\left( \vec{p},\vec{q}\right) =-\frac{1}{2}\left\vert \vec{q}%
\times \vec{p}\,\right\vert ^{2}-\frac{1}{2}\kappa \left(
xq_{x}+yq_{y}\right) .
\end{equation*}
Meanwhile, $\sigma \left( \varrho \right) \equiv \kappa \left( \varrho
\varsigma _{z}+\varsigma _{z}\varrho \right) -\left\langle \varrho ,2\kappa
\varsigma _{z}\right\rangle \varrho $ and so 
\begin{equation*}
\left\langle \sigma \left( \varrho \right) ,Q\right\rangle =-\kappa
zxq_{x}-zyq_{y}+\kappa \left( 1-z^{2}\right) q_{z}.
\end{equation*}

With the customary abuse of notation, we write $\mathsf{S}\left( t,\varrho
\right) \equiv \mathsf{S}\left( t,x,y,z\right) $. The It\^{o} correction
term, $\frac{1}{2}\left\langle \sigma \left( \varrho \right) \otimes \sigma
\left( \varrho \right) ,\delta \otimes \delta \mathsf{S}\right\rangle $, in
the HJB equation is then given by $\left( \text{with }\mathsf{S}_{xy}=\frac{%
\partial \mathsf{S}}{\partial x\partial y}\text{, etc.}\right) $%
\begin{equation*}
\frac{\kappa ^{2}}{2}\left( 
\begin{array}{ccc}
-zx, & -zy, & 1-z^{2}
\end{array}
\right) \left( 
\begin{array}{ccc}
\mathsf{S}_{xx} & \mathsf{S}_{xy} & \mathsf{S}_{xz} \\ 
\mathsf{S}_{yx} & \mathsf{S}_{yy} & \mathsf{S}_{yz} \\ 
\mathsf{S}_{zx} & \mathsf{S}_{zy} & \mathsf{S}_{zz}
\end{array}
\right) \left( 
\begin{array}{c}
-zx \\ 
-zy \\ 
1-z^{2}
\end{array}
\right) .
\end{equation*}
Putting everything together, we find that the Hamilton-Jacobi-Bellman
equation is 
\begin{eqnarray*}
0 &=&\frac{\partial \mathsf{S}}{\partial t}-\frac{1}{2}\left\vert \vec{q}%
\times \vec{\nabla}\mathsf{S}\right\vert ^{2}-\frac{1}{2}\kappa \left( x%
\frac{\partial \mathsf{S}}{\partial x}+y\frac{\partial \mathsf{S}}{\partial y%
}\right) \\
&&+\frac{\kappa ^{2}}{2}\left( x^{2}z^{2}\frac{\partial ^{2}\mathsf{S}}{%
\partial x^{2}}+y^{2}z^{2}\frac{\partial ^{2}\mathsf{S}}{\partial y^{2}}%
+\left( 1-z^{2}\right) ^{2}\frac{\partial ^{2}\mathsf{S}}{\partial z^{2}}%
\right. \\
&&+\left. xyz^{2}\frac{\partial ^{2}\mathsf{S}}{\partial x\partial y}%
-xz\left( 1-z^{2}\right) \frac{\partial ^{2}\mathsf{S}}{\partial x\partial z}%
-yz\left( 1-z^{2}\right) \frac{\partial ^{2}\mathsf{S}}{\partial y\partial z}%
\right) .
\end{eqnarray*}

\section{Discussion}

In our analysis we have sought to think of the quantum state of a controlled
system (that is, its von Neumann density matrix) in the same spirit as
classical control engineers think about the state of the system. The
advantage of this is that all the quantum features of the problem are
essentially tied up in the state: once the measurements have been performed
the information obtained can be treated as essentially classical, as can the
problem of using this information to control the system in an optimal
manner. The disadvantage is that we have to deal with a stochastic
differential equation on the infinite dimensional space of quantum states.
Nevertheless, the Bellman principle can then be applied in much the same
spirit as for classical states and we are able to derive the corresponding
Hamilton-Jacobi-Bellman theory for a wider class of cost functionals than
traditionally considered in the literature. When restricted to a
finite-dimensional representation of the state (on the Bloch sphere for the
qubit) with the cost being a quantum expectation, we recover the class of
Bellman equations encountered as standard in quantum feedback control.

\bigskip

\textbf{Acknowledgment}

We would like to thank Luc Bouten, Ramon van Handel, Hideo Mabuchi, Aubrey
Truman for useful discussions. J.G. would like to acknowledge the support of
EPSRC research grant GR/R78404/01, and V.P.B. acknowledges EEC support
through the ATESIT project IST-2000-29681 and the RTN network
QP\&Applications.

\end{document}